\documentclass[conference]{IEEEtran}
\IEEEoverridecommandlockouts
\usepackage{cite}
\usepackage{amsmath,amssymb,amsfonts}
\usepackage{algorithmic}
\usepackage{algorithm}
\usepackage{graphicx}
\usepackage{textcomp}
\usepackage{xcolor}
\usepackage{hyperref}
\usepackage{url}
\usepackage[subrefformat=parens]{subcaption}
\usepackage{siunitx}
\usepackage{dsfont}

\newcommand{\figref}[1]{Fig.\,\ref{#1}}
\newcommand{\Figref}[1]{Figure\,\ref{#1}}

\newcommand{\eref}[1]{Eq.\,(\ref{#1})}
\newcommand{\secref}[1]{Sec.\,\ref{#1}}
\newcommand{\ie}{\emph{i.e.}}

\newcommand{\mI}{{\mathds{1}}}

\begin{document}

\title{
Probabilistic Reachability Analysis of Multi-scale Voltage Dynamics Using Reinforcement Learning
\thanks{This work was supported in part by Grant-in-Aid for Scientific Research (KAKENHI) from the Japan Society for Promotion of Science (\#23K13354).
The Fraunhofer IEE work was supported by the Federal Ministry for Economic Affairs and Energy and the Projekträger Jülich as part of the (eKI4DS) project (FKZ 03EI1092D) and (LISA-FKZ03EI4059A). The authors are solely responsible for the content of this publication and it does not necessarily reflect the opinion of the entire project consortium.}
}

\author{
%%%%%%%%%%%%%%%%%%%%%
% \IEEEauthorblockN{Naoki Hashima}
% \IEEEauthorblockA{\textit{Dept. of Electrical Materials and Engineering} \\
% \textit{University of Hyogo}\\
% Himeji, Japan \\
% er24y025@guh.u-hyogo.ac.jp}
% \and
% \IEEEauthorblockN{Hikaru Hoshino}
% \IEEEauthorblockA{\textit{Dept. of Electrical Materials and Engineering} \\
% \textit{University of Hyogo}\\
% Himeji, Japan \\
% hoshino@eng.u-hyogo.ac.jp}
% \and
% \IEEEauthorblockN{Luis David Pab\'{o}n Ospina} 
% \IEEEauthorblockA{\textit{Grid Control and Grid Dynamics} \\
% \textit{Fraunhofer IEE}\\
% Kassel, Germany \\
% luis.david.pabon.ospina@iee.fraunhofer.de}
% \and
% \IEEEauthorblockN{Eiko Furutani}
% \IEEEauthorblockA{\textit{Dept. of Electrical Materials and Engineering} \\
% \textit{University of Hyogo}\\
% Himeji, Japan \\
% furutani@eng.u-hyogo.ac.jp}
%%%%%%%%%%%
\IEEEauthorblockN{Naoki Hashima${}^\dagger$, Hikaru Hoshino${}^\dagger$, Luis David Pab\'on Ospina${}^\ddagger$, Eiko Furutani${}^\dagger$}\\[-2mm]
\begin{minipage}[t]{.56\linewidth}
\centering
${}^\dagger$\textit{Department of Electrical Materials and Engineering} \\
\textit{University of Hyogo}\\
Himeji, Japan \\
\{er25o015@guh, hoshino@eng, furutani@eng\}.u-hyogo.ac.jp
\end{minipage}\hfill
\begin{minipage}[t]{.42\linewidth}\centering
${}^\ddagger$\textit{Grid Control and Grid Dynamics}\\
\textit{Fraunhofer IEE}\\
Kassel, Germany\\
luis.david.pabon.ospina@iee.fraunhofer.de
\end{minipage} 
%%%%%%%%%%%%%%%%%%%%%%%%%%%%%%%%%%%%
% \begin{minipage}[t]{.48\linewidth}\centering
% {\fontsize{11pt}{11pt}\selectfont Naoki Hashima}\\
% \textit{Dept. of Electrical Materials and Engineering}\\
% \textit{University of Hyogo}\\
% Himeji, Japan\\
% er25o015@guh.u-hyogo.ac.jp
% \end{minipage}\hfill
% \begin{minipage}[t]{.48\linewidth}\centering
% {\fontsize{11pt}{11pt}\selectfont Hikaru Hoshino}\\
% \textit{Dept. of Electrical Materials and Engineering}\\
% \textit{University of Hyogo}\\
% Himeji, Japan\\
% {hoshino@eng.u-hyogo.ac.jp}
% \end{minipage}\\[0.9em]
% \vspace{0.8em}
% \begin{minipage}[t]{.48\linewidth}\centering
% {\fontsize{11pt}{11pt}\selectfont Luis David Pab\'on Ospina}\\
% \textit{Grid Control and Grid Dynamics}\\
% \textit{Fraunhofer IEE}\\
% Kassel, Germany\\
% luis.david.pabon.ospina@iee.fraunhofer.de
% \end{minipage}\hfill
% \begin{minipage}[t]{.48\linewidth}\centering
% {\fontsize{11pt}{11pt}\selectfont Eiko Furutani}\\
% \textit{Dept. of Electrical Materials and Engineering}\\
% \textit{University of Hyogo}\\
% Himeji, Japan\\
% furutani@eng.u-hyogo.ac.jp
% \end{minipage}
}

\maketitle

\begin{abstract}
Voltage stability in modern power systems involves coupled dynamics across multiple time scales. Conventional methods based on time-scale separation or static stability margins may overlook instabilities caused by the coupling of slow and fast transients. Uncertainty in operating conditions further complicates stability assessment, and high computational cost of Monte Carlo simulations limit its applicability to multi-scale dynamics. This paper presents a deep reinforcement learning-based framework for probabilistic reachability analysis of multi-scale voltage dynamics. By formulating each instability mechanism as a distinct absorbing state and introducing a multi-critic architecture for mechanism-specific learning, the proposed method enables consistent learning of risk probabilities associated with multiple instability types within a unified framework. The approach is demonstrated on a four-bus system with load tap changers and over-excitation limiters, illustrating effectiveness of the proposed learning-based reachability analysis in identifying and quantifying the mechanisms leading to voltage collapse.
\end{abstract}

\begin{IEEEkeywords}
Voltage stability assessment, Deep RL,
Probabilistic stability analysis, Singular perturbation 
\end{IEEEkeywords}

%%%%%%%%%%%%%%%%%%%%%%%%%%%%%%%%%%%%%%%%%%%%%%%%%%%%%%%%%%
\section{Introduction}
%\rev{Luis: I will include all my proposals in red. Please feel free to accept/reject/modify them}

In power systems, a wide range of dynamic phenomena emerge, each characterized by distinct physical mechanisms and time scales. 
To improve tractability and computational efficiency, time-scale separation is commonly applied in power system stability analysis~\cite{Machowski2008}: studies focusing on fast dynamics often disregard slow ones, and those examining slow dynamics tend to overlook fast ones.
While this simplification has long supported practical stability assessment, its validity becomes questionable as modern power systems integrate more renewable and inverter-based resources~\cite{PabonOspina2023,PabonOspina2025PowerTech,PabonOspina2025IBRs}.
In particular, it is demonstrated in \cite{PabonOspina2025PowerTech,PabonOspina2025IBRs} that conventional static-margin methods, quasi-static analysis or even time-domain simulation can fail to detect voltage instabilities where the evolution of long-term dynamics triggers instability in short-term dynamics.

Beyond modeling fidelity, another key challenge arises from the uncertainty inherent in operating conditions.
Fluctuations in load demand, renewable generation, and system parameters introduce stochasticity that directly impacts system stability.
Monte Carlo simulation is widely used to capture such uncertainty, estimating probabilities of voltage collapse over many randomized scenarios. %~\cite{Almeida2013,Hasan2019}.
However, due to their high computational cost, such methods typically rely on quasi-static or reduced-order models~\cite{Hasan2019}.  
Yet, as noted above, such simplifications may fail to capture critical instability mechanisms arising from multi-scale dynamics. %, leading to potential underestimation of risk probabilities.

To address these challenges, this paper uses the concept of probabilistic reachability~\cite{Abate2008,Schmid2023} as a basis for analyzing multi-scale voltage dynamics under uncertainty.
This concept provides a principled way to quantify the risk probability that system trajectories reach a defined unsafe set, explicitly accounting for the stochastic and nonlinear nature of the dynamics. 
However, reachability computation in nonlinear systems usually requires solving a dynamic programming equation, which becomes computationally intractable for high-dimensional models due to the curse of dimensionality~\cite{Ganai2024}.
To overcome this barrier, we use deep Reinforcement Learning (RL) as a scalable approximation method through data-driven value function learning~\cite{Ganai2024}.
This approach not only retains the theoretical foundation of reachability analysis but also enables efficient exploration of preventive control actions and real-time stability assessment once training is completed.

The contribution of this paper is two-fold. 
First, we present an RL-based probabilistic reachability analysis framework tailored for stability assessment of multi-scale voltage dynamics.
Based on the second author's previous formulation of RL-based probabilistic reachability~\cite{HoshinoACC2024}, this work extends the approach to account for multiple instability mechanisms that may concurrently arise in power systems. 
Specifically, the proposed framework enables learning of mechanism-specific risk probabilities, quantifying the contribution of each instability mechanism to the overall voltage-collapse risk.
This extension provides interpretable insight into the dominant mechanisms of voltage collapse and facilitates informed decision making under uncertainty.

Second, the effectiveness of the proposed framework is demonstrated using a four-bus test system originally introduced by \cite{VanCutsem1998}, which serves as a concept-friendly yet still representative example of multi-scale voltage instability.
The system includes both long-term dynamics induced by Load Tap Changers (LTCs) and Over-Excitation Limiters (OXLs), and short-term dynamics associated with Automatic Voltage Regulators (AVRs), generator responses, and motor behavior.
Numerical experiments show that the proposed method effectively captures the coupled long- and short-term dynamics and demonstrates its applicability to multi-scale voltage stability analysis under uncertainty.

Deterministic reachability analysis has been previously applied to power system analysis, including voltage stability~\cite{Susuki2007}, transient stability~\cite{Jin2010,Susuki2011}, and microgrid control problems~\cite{Li2018}, as reviewed in~\cite{Zhang2020}.
In contrast, to the best of our knowledge, this study is the first to apply probabilistic reachability to power system analysis, enabling direct quantification of risk under stochastic operating conditions.

The rest of this paper is organized as follows. 
In \secref{sec:multi-scale}, multi-scale voltage dynamics studied in this paper are introduced. 
In \secref{sec:method}, we describe the proposed framework.  
In \secref{sec:simulation}, numerical experiments are given followed by conclusion in \secref{sec:conclusion}.

%%%%%%%%%%%%%%%%%%%%%%%%%%%%%%%%%%%%%%%%%%%%%%%%%%%%%%%%%%%%%
\section{Multi-scale Voltage Dynamics} \label{sec:multi-scale}

This section reviews the multi-scale voltage dynamics considered in this study.
In \secref{sec:scale_separation}, time-scale separation and quasi-static approximation used in standard voltage stability studies are explained. 
In \secref{sec:instability}, we describe the multi-scale instability mechanisms that emerge when this scale separation breaks down.

%%%%%%%%%%%
\subsection{Time-scale Separation and Quasi-static Approximation} \label{sec:scale_separation}

Power system dynamics involve a variety of processes evolving over different time scales. 
While discontinuous dynamics such as LTCs will be considered later, we focus here on the continuous-time dynamics for simplicity of exposition:
\begin{align}
    \dot{x} & = f(x,y), \\
    \epsilon \dot{y} & = g(x,y)
\end{align}
where $x$ stands for the slow variables, $y$ for the fast ones, and the dot symbol represents their time derivative. 
$\epsilon \ll 1$ is the ratio of their time constants.
The quasi-static approximation, commonly adopted in long-term voltage stability studies, assumes that the fast subsystem remains at its equilibrium for any given $x$, \ie,
\begin{align}
    g(x,y) = 0.   
\end{align}
By letting the equilibrium solution of this algebraic equation be described as $y=\phi(x)$, the mapping $\phi$ defines the fast subsystem equilibrium manifold parameterized by the slow variables, and the system evolution is governed only by the slow dynamics:
\begin{align}
  \dot{x} = f(x, \phi(x)).
\end{align}
This approximation is valid only as long as the equilibrium 
$\phi(x)$ exists and remains stable.
In real power systems, slow devices continuously modify operating conditions, and these changes may gradually push the fast subsystem toward a loss of equilibrium or a change in its stability type.
%Once this occurs, the system can no longer follow the quasi-static trajectory, and a fast instability develops even though static indicators still predict feasibility.

%%%%%%%%%%%
\subsection{Multi-scale Instability Mechanisms} \label{sec:instability}

\begin{figure}
    \centering
    \includegraphics[width=0.8\linewidth]{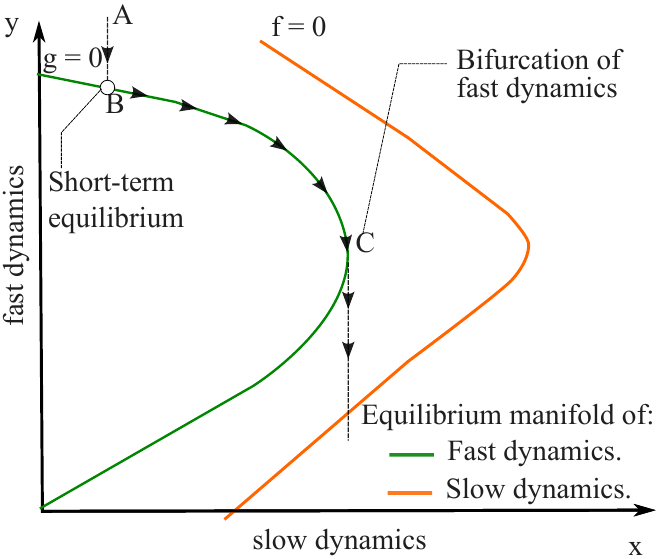}
    \caption{Loss of short-term equilibrium due to slow dynamics  (modified from \cite{IEEE2002PES_TR9})}
    \label{fig:multi-scale_instability}
\end{figure}

The breakdown of time-scale separation can be illustrated by using the slow and fast equilibrium manifolds (defined by $f(x,y)=0$ and $g(x,y)=0$, respectively) as shown in \figref{fig:multi-scale_instability}.
Following a disturbance, the system rapidly moves from its pre-disturbance state (point A) to a short-term equilibrium (point B) along the fast dynamics, while the slow variables remain nearly constant.
As the slow dynamics subsequently evolve, they drive the operating point toward the slow equilibrium manifold.
However, when the slow and fast manifolds do not intersect, the fast subsystem loses its equilibrium before reaching the long-term steady state (point C), resulting in a loss of short-term equilibrium.
This mechanism represents one of the typical forms of multi-scale  instability, and three types of instability have been identified in~\cite{VanCutsem1998}:
\begin{itemize}
  \item S-LT1: Loss of short-term equilibrium due to slow variable evolution. 
  \item S-LT2: Shrinkage of the region of attraction of the short-term equilibrium, leading to divergence of trajectories even before equilibrium is lost.
  \item S-LT3: Oscillatory instability of fast dynamics, where the slow evolution induces a Hopf bifurcation in the short-term subsystem.
\end{itemize}
Among these, the case illustrated in \figref{fig:multi-scale_instability} corresponds to S-LT1, which is most commonly observed in classical voltage-stability problems. 
S-LT2 mechanisms have been reported in transient-stability contexts, while more recently, inverter-driven oscillations corresponding to S-LT3 have been reported in~\cite{PabonOspina2023,PabonOspina2025IBRs}. 
In this study, we primarily focus on S-LT1 instability %which represents the most typical form of multi-scale voltage instability. 
and %Within this category, we 
further distinguish multiple mechanisms that lead to the loss of short-term equilibrium in generator or motor dynamics, and estimate their respective failure probabilities using the proposed probabilistic reachability framework.
In principle, the same framework can also be extended to analyze other types of instabilities (including S-LT2 and S-LT3), although their detailed investigation is left for future work.

%%%%%%%%%%%%%%%%%%%%%%%%%%%%%%%%%%%%%%%%%%%%%%%%%%%%%%%%%%
\section{Proposed Analysis Framework} \label{sec:method}

This section presents the proposed probabilistic reachability analysis framework for multi-scale voltage dynamics.
Building on the RL-based probabilistic reachability formulation introduced in \secref{sec:reachability}, we extend the approach to handle multiple instability mechanisms in \secref{sec:proposed_framework}.

%%%%%%%%%%%%%%%%
\subsection{RL-Based Probabilistic Reachability} \label{sec:reachability}

This study considers the invariance-type (or avoidance-type) reachability problem that aims to quantify the likelihood that system trajectories remain within a predefined safe set over a finite horizon.
Equivalently, the complement of this probability represents the risk probability.
The formulation is based on our previous work~\cite{HoshinoACC2024} with a focus on the model-free setting.
%This study considers the avoidance-type reachability problem, which aims to estimate the probability of avoiding an unsafe set, based on the RL-based formulation presented in~\cite{HoshinoACC2024} with a focus on the model-free setting. 

Let $\mathcal{Z} \subset \mathbb{R}^n$ be the state space, and 
$\mathcal{U} \subset \mathbb{R}^m$ be the control space. 
Consider a discrete-time stochastic system evolving at time steps $k = 0, 1, 2, \dots$, defined by the transition kernel
\begin{align}
    T(\mathrm{d}z'|z, u)
    := \Pr[z_{k+1} \in \mathrm{d}z' \mid z_k = z,\, u_k = u],
    \label{eq:transition_kernel}
\end{align}
which specifies the conditional probability distribution of the next state $z_{k+1}$ given $(z_k, u_k)$. 
A stationary Markov policy $\mu:\mathcal{Z}\!\to\!\mathcal{U}$
induces a controlled Markov process $\{z_k\}_{k\ge0}$ satisfying
$z_{k+1}\sim T(\,\cdot\,|z_k,\mu(z_k))$.
Let $\mathcal{C}\subset\mathcal{Z}$ be the safe set,
$\mathcal{C}^{\mathrm{c}}$ be its complement (unsafe set). 
Then, the \emph{safety probability} (complement of risk probability) under a policy~$\mu$ for a time horizon~$\tau>0$, starting from an initial state~$z_0$, is defined as
\begin{align}
    \Psi^{\mu}(\tau, z_0)
    := \Pr\!\left[z_k\!\in\!\mathcal{C},\,\forall k\!\le\!N(\tau)
        \,\middle|\,z_0,\,\mu\right],
    \label{eq:avoidance_prob}
\end{align}
where $N(\tau):=\lfloor\tau/\Delta t\rfloor$ is the largest integer not greater than $\tau / \Delta t$, representing the number of discrete time steps within the horizon~$\tau \in \mathbb{R}$.
The goal of the probabilistic reachability analysis is to compute the
maximal safety probability 
\begin{align}
    \Psi^{*}(\tau, z_0)
    = \sup_{\mu\in\mathcal{U}} \Psi^{\mu}(\tau, z_0),
    \label{eq:maxsafeprob}
\end{align}
which quantifies the highest achievable probability of avoiding
$\mathcal{C}^{\mathrm{c}}$ during the horizon~$\tau$. 

% In general, computing the maximal safety probability~\eqref{eq:maxsafeprob} requires solving a dynamic programming equation~\cite{Abate2008,Schmid2023}, which becomes computationally intractable for high-dimensional nonlinear systems.
% To overcome this issue, our previous work~\cite{HoshinoACC2024} proposed an RL-based approximation framework that learns the maximal safety probability. 

To enable RL-based computation of the maximal safety probability, it is necessary to augment the system state and define a suitable reward structure, which are described below. 

%%%%%%%%%%%%%%%%%%%%
\subsubsection{Augmented-State Formulation} 

The probability in~\eref{eq:avoidance_prob} can be expressed as a multiplicative cost over time:
\begin{align}
    \Psi^{\mu}(\tau, z_0)
    = \mathbb{E}\!\left[
        \prod_{k=0}^{N(\tau)} \mI_{\mathcal{C}}(z_k)
        \,\middle|\,z_0,\,\mu
    \right], 
    \label{eq:multiplicative_cost}
\end{align}
where $\mI_{\mathcal{C}}(z)$ stands for the indicator function, which takes the value~1 if $z \in \mathcal{C}$ and~0 otherwise, and $\mathbb{E}[\cdot]$ stands for the expectation over the stochastic transitions.
This multiplicative structure is incompatible with standard RL,
which assumes additive returns.
To overcome this, we introduce an augmented state that
explicitly tracks the remaining time before the horizon~$\tau$ is reached:
\begin{align}
    h_{k+1} = h_k - \Delta t, \qquad h_0 = \tau,
\end{align}
and define
\begin{align}
    s_k = [\,h_k,\, z_k^{\top}\,]^{\top}
    \in \mathcal{S} := \mathbb{R}\times\mathcal{Z}.
\end{align}
The set of absorbing states is then defined as
\begin{align}
    \mathcal{S}_{\mathrm{abs}}
    := \{[h,z]^{\top}\!\in\!\mathcal{S}\mid
        h<0 \ \text{or}\ z \!\in\!\mathcal{C}^\mathrm{c}\},
\end{align}
so that once $s_k$ enters $\mathcal{S}_{\mathrm{abs}}$, it remains there.
The augmented transition kernel is then given by
\begin{align}
    \tilde{T}(s'|s,u) :=
    \begin{cases}
        \delta(s'\!-\![h-\Delta t,\,z']^{\top})
            T(z'|z,u), & s\notin\mathcal{S}_{\mathrm{abs}},\\[1.5mm]
        \delta(s'\!-\!s), & s\in\mathcal{S}_{\mathrm{abs}},
    \end{cases}
\end{align}
where $\delta(\cdot)$ denotes the Dirac measure.

%%%%%%%%%%%%%%%%%%%
\subsubsection{Reward Definition} 

The RL reward is defined by
\begin{align}
    r(s) := \mI_{[0,\Delta t)}(h)\,\mI_{\mathcal{S}\setminus\mathcal{S}_{\mathrm{abs}}}(s),
    \label{eq:reward_def}
\end{align}
which yields a unit reward when the remaining time $h$ lies within $[0,\Delta t)$ and the system has not entered the absorbing set $\mathcal{S}_{\mathrm{abs}}$. 
Then, for any policy~$\mu$, the value function in the augmented space is given by
\begin{align}
    v^{\mu}(s_0)
    := \mathbb{E}\!\left[
        \sum_{k=0}^{\infty} r(s_k)
        \,\middle|\,s_0,\,\mu
    \right].
    \label{eq:value_def}
\end{align}
It is proved in~\cite{HoshinoACC2024} that $v^{\mu}(s_0)=\Psi^{\mu}(\tau,z_0)$ with $s_0 = [\tau, z_0^\top]^\top$ and therefore
\begin{align}
    v^{*}(s_0)
    = \sup_{\mu} v^{\mu}(s_0)
    = \Psi^{*}(\tau,z_0).
\end{align}

This equivalence reformulates the safety probability calculation as a
standard expected-return maximization problem, allowing direct use of
off-the-shelf deep RL algorithms (\emph{e.g.}, DQN, DDPG, TD3). 
In these model-free algorithms, explicit knowledge of the transition kernel is not required. 
It suffices to perform time-domain simulations of state transitions, and the underlying dynamics may include discontinuities.

%%%%%%%%%%%%%%%%%%
\subsection{Mechanism-specific Analysis Framework} \label{sec:proposed_framework}

This subsection presents an extension of the above RL-based probabilistic reachability to analyze multi-scale voltage instability.
The proposed approach combines trajectory-level exploration of long-term dynamics with detection and attribution of short-term instability events.

\subsubsection{Instability Detection}

In multi-scale voltage dynamics, short-term instability may arise when the fast subsystem loses its equilibrium or undergoes a change in stability, corresponding to point~C in \figref{fig:multi-scale_instability}.
To identify such events along the trajectories generated along the RL training, several detection criteria can be employed.
For example, full-scale time-domain simulations can be used to monitor key state variables: large rotor-angle deviations approaching $360^{\circ}$ indicate pole slip and loss of synchronism in synchronous machines, whereas very low or zero motor speeds indicate motor instability.
Long-term dynamic simulations combined with small-signal analysis can be used to track the eigenvalue movement of the fast subsystem to detect bifurcation points.
%For example, a time-domain criterion can monitor abrupt variations in fast state variables such as rotor angle, internal voltage, or motor speed, while a small-signal criterion can track the eigenvalue movement of the fast subsystem to detect bifurcation points. 
%\rev{Besides abrupt variations in fast state variables, monitoring state values can also be an option. For example, large angle deviations close to $360^{\circ}$ are indicative of pole slip and loss of synchronism in synchronous machines. Zero or unusually low motor speeds are indicative of motor stalling.} 
Other mechanisms, such as shrinkage of the stability region, can also be incorporated depending on the system model and available data.
When any of these conditions are met, the corresponding state $z$ is regarded as having entered the unsafe set~$\mathcal{C}^{\mathrm{c}}$.
Within the above RL-based probabilistic reachability analysis, these unsafe regions are consistently represented in the state space, and the evolution of the slow dynamics can be analyzed to understand how the system approaches and eventually enters them.
This analysis can be interpreted as a backward reachability process from point~C to point~B in \figref{fig:multi-scale_instability}, identifying the regions in the state space that can lead to instability under the evolution of long-term dynamics.

%%%%%%%%%%%%%%%%%%%%
\subsubsection{Mechanism-Specific Learning}

Based on the instability detection described above, we propose a multi-critic RL architecture that enables mechanism-specific learning of risk probabilities. 
Let $\mathcal{M}=\{1,\ldots,M\}$ be the indices for distinct instability mechanisms (e.g., generator- and motor-induced S-LT1 instabilities, or other S-LT2 or S-LT3 instabilities).
We define disjoint absorbing sets
\begin{align}
    \{\mathcal{S}_{\mathrm{abs}}^{(m)}\}_{m\in\mathcal{M}}, \quad
    \mathcal{S}_{\mathrm{abs}}^{(m)}
    = \{ [h,z]^{\top} \in \mathcal{S} \mid z \in \mathcal{C}_{m}^{\mathrm{c}} \},
    \label{eq:s_abs_m}
\end{align}
where $\mathcal{C}_{m}^{\mathrm{c}}$ represents the unsafe set in the state space
corresponding to ``failure due to mechanism~$m$."
The overall absorbing set is then given by
\begin{align}
    \mathcal{S}_{\mathrm{abs}}
    = \{ [h,z]^{\top} \in \mathcal{S} \mid h < 0 \}
      \cup \bigcup_{m\in\mathcal{M}} \mathcal{S}_{\mathrm{abs}}^{(m)},
    \label{eq:s_abs_total}
\end{align}
which coincides with the unsafe set introduced earlier. %, and terminal labels $m$ can be identified by rule-based post-event classification. 
Based on this decomposition, a multi-dimensional reward $r^\mathcal{M} := [r^{(1)}, r^{(2)}, \dots, r^{(M)}]$ can be defined by
\begin{align}
r^{(m)}(s) := \mI_{[0,\Delta t)}(h)\,\mI_{\mathcal{S}\setminus\mathcal{S}_{\mathrm{abs}}^{(m)}}(s),
\quad m\in\mathcal{M},
\label{eq:multi_reward}
\end{align}
which yields a unit reward when the remaining time $h$ lies
within $[0, \Delta t)$ and the trajectory has not entered
$\mathcal{S}_{\mathrm{abs}}^{(m)}$. 
The $m\text{-th}$ critic $Q^{(m)}_{\theta_m}(s,a)$ is then trained via the standard TD update using
$r^{(m)}(s)$ as its reward, while a shared actor $\mu_\psi(s)$ is optimized to
maximize the minimum of these critic values:
\begin{align}
J_{\mathrm{actor}}(\psi)
= \mathbb{E}
\left[\min_{m\in\mathcal{M}}
Q^{(m)}_{\theta_m}\big(s,\mu_\psi(s)\big)\right].
\label{eq:actor_update}
\end{align}
Here, $J_{\mathrm{actor}}(\psi)$ represents the expected value of the worst-case critic, which serves as the optimization objective for updating the actor parameters~$\psi$.
This structure %of one actor with multiple critics 
enables mechanism-specific estimation of risk probabilities and unified policy learning %that improves the worst-case performance 
across multiple instability mechanisms.

%%%%%%%%%%%%%%%%%%%%%%%
\section{Numerical Experiments}
\label{sec:simulation}

\begin{figure}[b!]
    \centering
    \includegraphics[width=0.9\linewidth]{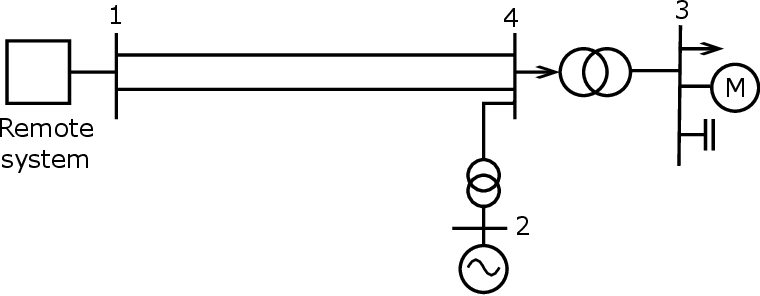}
    \caption{One-line diagram of the example system~\cite{VanCutsem1998}}
    \label{fig:test-system}
\end{figure}

\begin{figure}[b!]
    \centering
    \includegraphics[width=1.0\linewidth]{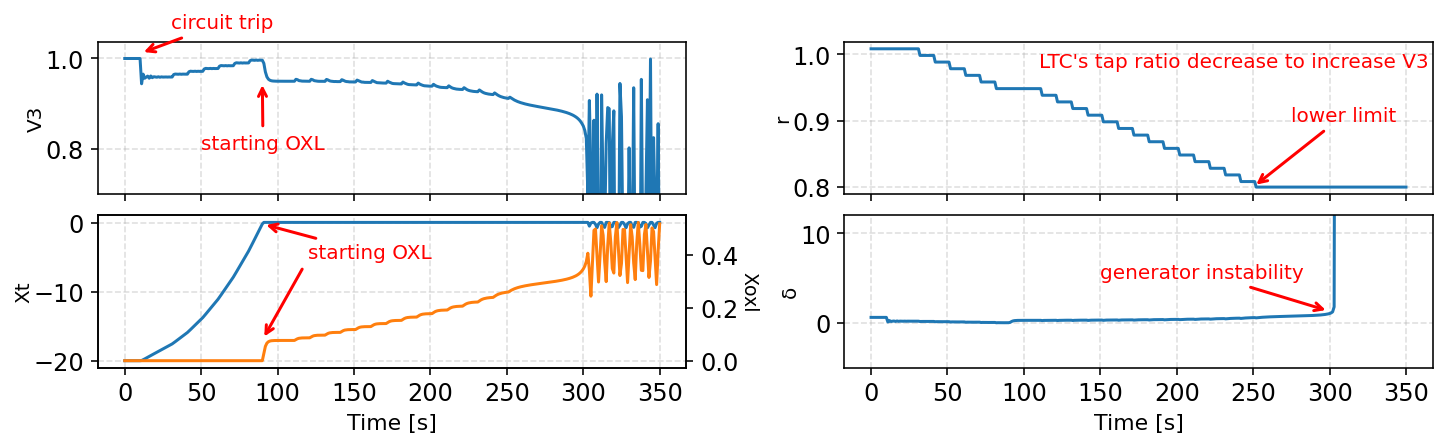}
    \caption{Time evolution of an S-LT1 type generator instability}
    \label{fig:S-LT1}
\end{figure}

The numerical experiments are conducted on the four-bus system introduced in~\cite[Sec.~6.3]{VanCutsem1998}.
The one-line diagram is shown in \figref{fig:test-system}. 
%The system consists of one synchronous generator (bus 1) supplying a remote load (bus 4) through two intermediate buses (2 and 3). \rev{Note: Fig 2 is not compiling, but as fat as I can see in figs/test\_system.svg, the SG is connected to bus 2 and not to bus 1. Bus 1 is an equivalent of the rest of the system. The load is connected to bus 3 and not to bus 4. Maybe the figure changed?} The transmission network includes series reactances $X_{14}=0.0277$ and a transformer with adjustable tap ratio $r$ controlled by a load-tap changer (LTC). The load at bus 4 is a composite of static and dynamic components, while the generator is equipped with an automatic voltage regulator (AVR) and an over-excitation limiter (OXL) \rev{this abbreviation was already defined before, it coan just be used in this case. Same for LTC}.
The load at bus 3 is fed through an LTC transformer. 
The total active power demand is \SI{1500}{MW} under nominal conditions, and most of the power is provided by the remote system through the tie line~1-4, while the remaining is supplied by the local generator at bus 2.
As a disturbance, the one-circuit tripping of the tie line is considered, which reduces the power transfer capability from the remote system and causes a voltage drop at the load bus.  
The LTC attempts to restore it toward a reference value by gradually decreasing the tap ratio, which acts as a slow control mechanism in the system. 
This configuration enables realistic interaction between fast transients such as  electromechanical dynamics of generators and motors, and internal voltage and AVR actions, and slow transients such as LTC tap movement and OXL activation. 
\Figref{fig:S-LT1} shows an exmaple time evolution of an S-LT1 type generator instability. % observed at the system. %when the generator at bus~2 operates at \SI{450}{MW} (\SI{90}{\%} of its rated power).
Following the disturbance at {$t=\SI{10}{s}$}, the LTC gradually lowers its tap ratio to restore the load-side voltage, increasing the reactive demand on the generator.
At around {$t=\SI{90}{s}$}, the excitation reaches the OXL limit, and the internal voltage starts to decline.
The system remains quasi-stable for a while, but at around {$t=\SI{300}{s}$} the generator loses equilibrium, resulting in voltage collapse.

%%%%%%%%%%%%%%%%%%%
%\subsection{Results}

The instability risk under uncertain operating conditions is next evaluated through the proposed method. 
The observable state in the training differs from the full set of simulation states and is defined as 
\begin{align}
 z = [V_4, \, E'_q, X_\mathrm{oxl}; P_\mathrm{g}, R_\mathrm{motor}]^\top,
\end{align}
where $V_4$ is the voltage at bus~4, $E'_q$ the generator internal voltage, and $X_\mathrm{oxl}$ the OXL state variable, which together represent the dynamic states of the system.
The last two variables, $P_\mathrm{g}$ and $R_\mathrm{motor}$, indicate operating conditions corresponding to the generator active power and motor load ratio, respectively. 
In the study, the nominal total demand is fixed at \SI{1500}{MW}, and 
the ratio between the motor and the standard exponential load components changes with the rated power of the motor kept constant. 
To emulate uncertainty in operating conditions, Gaussian perturbations are added to the total demand and the motor ratio, with standard deviations of \SI{5}{MW} and $0.05$, respectively.
%To emulate uncertainty in operating conditions, a Gaussian perturbation with a standard deviation of {\SI{5}{MW}} is added. Similarly, random noise is applied to the actual motor ratio with a standard deviation of {$0.05$} to represent uncertainty in $R_\mathrm{motor}$. 
The corrective action is defined as an incremental adjustment of the LTC reference~$V_3^{\mathrm{ref}}$.
At each time step, the actor output in~$[-1,1]$ is scaled by~0.1 and added to the previous reference value, with~$V_3^{\mathrm{ref}}$ clipped within~$0.9$ to $1.1$~pu.
For the training, the TD3 algorithm is used with {three}-layer actor and critic networks, each consisting of {64} units per layer.
Adam learning rates are set to {$10^{-5}$} for the actor and $10^{-4}$ for the critic. 
%none motor simulation learning rate actor: $10^{-5}$, critic:$10^{-4}$,
%considering motor simulation actor:$10^{-6}$ critic: $10^{-4}$

\begin{figure}[!t]
  \begin{minipage}[b]{0.495\linewidth}
    \centering
    \includegraphics[width=\linewidth]{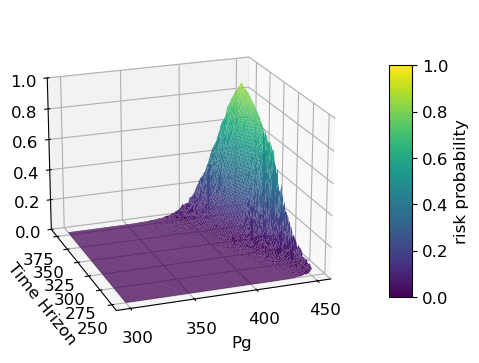}
    \subcaption{Without corrective action}
  \end{minipage}
  \begin{minipage}[b]{0.495\linewidth}
    \centering
    \includegraphics[width=\linewidth]{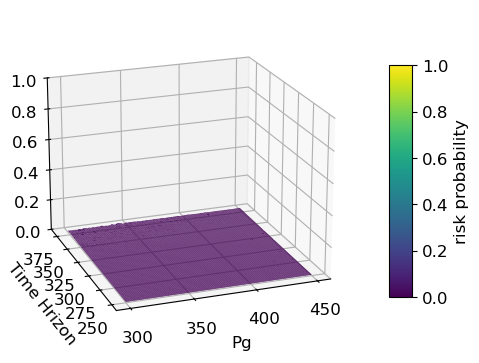}
    \subcaption{With corrective action}
  \end{minipage}
  \caption{Learned risk probability with $R_\mathrm{motor}=0$} \label{fig:non_motor}
  \begin{minipage}[b]{0.495\linewidth}
    \centering
    \includegraphics[width=\linewidth]{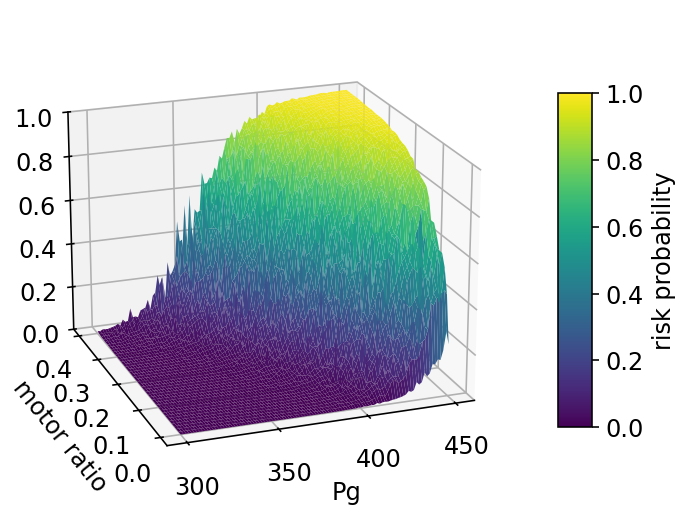}
    \subcaption{Generator instability}
  \end{minipage}
  \begin{minipage}[b]{0.495\linewidth}
    \centering
    \includegraphics[width=\linewidth]{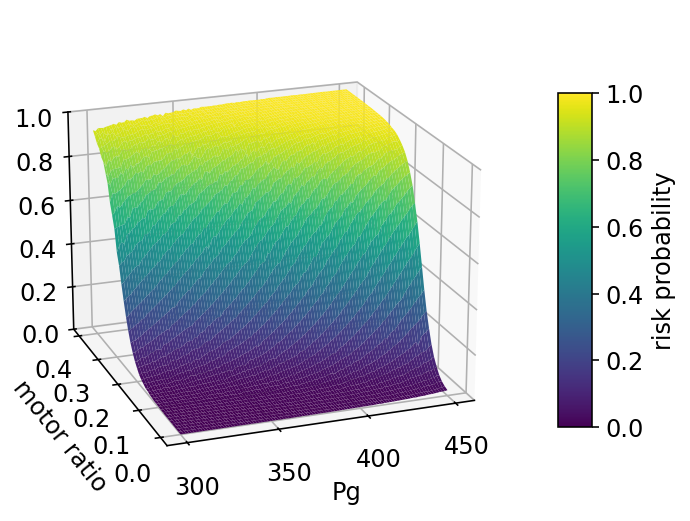}
    \subcaption{Motor instability}
  \end{minipage}
  \caption{Learned risk probability as a function of $P_\mathrm{g}$ and $R_\mathrm{motor}$} \label{fig:gen_motor}
\end{figure}

\Figref{fig:non_motor} shows the learned risk probability (complement of safety probability) in the case without motor dynamics. 
The risk is evaluated as a function of the horizon $\tau$ and generator active power~$P_\mathrm{g}$.
Since the considered instability is triggered by long-term dynamics, the risk probability increases with longer horizons, indicating that the likelihood of losing short-term equilibrium grows as the slow variables evolve.
Higher~$P_\mathrm{g}$ leads to a greater instability risk, consistent with the conventional understanding that increasing mechanical power reduces the stability margin.
Panel (a) corresponds to the nominal case without corrective action, whereas panel (b) shows that the LTC reference adjustment effectively reduces the risk probability across horizons.
These trends show a good correlation with the expected behavior from a power system perspective, but such quantification has been difficult in complex systems without the proposed method. 
%\rev{NOTE: This is just a note according to my understanding, please feel free to rephrase it. I think that Fig.3 already shows a good correlation between the method and the expected behavior from a power system perspective. It is clear (even from a simple equal-area criterion perspective) that when the mechanical power (Pm) of the synchronous machine is increased, the stability margin decreases. This is quantified by the method as the estimated probability Q that decreases as Pm increases. Without the proposed method, such quantification is difficult in complex systems.}

\Figref{fig:gen_motor} extends the analysis by incorporating motor dynamics and shows the risk probability learned  without corrective action. 
Panels (a) and (b) show the probabilities of generator- and motor-induced instabilities, respectively, as a function of $P_\mathrm{g}$ and $R_\mathrm{motor}$.
As the generator output~$P_\mathrm{g}$ increases, the risk of generator instability rises due to a reduced stability margin, whereas the risk of motor instability becomes prominent at higher motor load ratios~$R_\mathrm{motor}$, reflecting the tendency of induction motors to stall under low-voltage conditions.
These results demonstrate that the proposed multi-critic learning framework successfully captures mechanism-specific behaviors, allowing quantitative differentiation of the dominant instability sources within a unified learning framework.

%%%%%%%%%%%%%%%%%%%%%%%%%%%%%%%%%%%%%%%%%%%
\section{Conclusions}
\label{sec:conclusion}

This paper proposed an RL-based probabilistic reachability framework for analyzing multi-scale voltage instability. 
By formulating instability mechanisms as absorbing states and introducing a
multi-critic architecture for mechanism-specific learning, the proposed framework enables consistent estimation of risk probabilities across multiple instability types.
The approach was validated on a four-bus system with LTC and OXL dynamics, demonstrating its capability to quantify and differentiate the mechanisms leading to voltage collapse.
Future work will extend the framework to larger systems including renewable resources and additional instability types.

\bibliographystyle{ieeetr}
\bibliography{ref}

% \begin{thebibliography}{00}
% \bibitem{b1} G. Eason, B. Noble, and I. N. Sneddon, ``On certain integrals of Lipschitz-Hankel type involving products of Bessel functions,'' Phil. Trans. Roy. Soc. London, vol. A247, pp. 529--551, April 1955.
% \bibitem{b2} J. Clerk Maxwell, A Treatise on Electricity and Magnetism, 3rd ed., vol. 2. Oxford: Clarendon, 1892, pp.68--73.
% \bibitem{b3} I. S. Jacobs and C. P. Bean, ``Fine particles, thin films and exchange anisotropy,'' in Magnetism, vol. III, G. T. Rado and H. Suhl, Eds. New York: Academic, 1963, pp. 271--350.
% \bibitem{b4} K. Elissa, ``Title of paper if known,'' unpublished.
% \bibitem{b5} R. Nicole, ``Title of paper with only first word capitalized,'' J. Name Stand. Abbrev., in press.
% \bibitem{b6} Y. Yorozu, M. Hirano, K. Oka, and Y. Tagawa, ``Electron spectroscopy studies on magneto-optical media and plastic substrate interface,'' IEEE Transl. J. Magn. Japan, vol. 2, pp. 740--741, August 1987 [Digests 9th Annual Conf. Magnetics Japan, p. 301, 1982].
% \bibitem{b7} M. Young, The Technical Writer's Handbook. Mill Valley, CA: University Science, 1989.
% \end{thebibliography}

\end{document}